# Dual-functional microwave photonic system for concurrent radar and secure communication via radar signal masking

Taixia Shi,[a,b,c] Fangzheng Zhang,[b] and Yang Chen[a,c,*]

[a] Shanghai Key Laboratory of Multidimensional Information Processing, East China Normal University, Shanghai, 200241, China
[b] Key Laboratory of Radar Imaging and Microwave Photonics, Ministry of Education, Nanjing University of Aeronautics and Astronautics, Nanjing 210016, China
[c] Engineering Center of SHMEC for Space Information and GNSS, East China Normal University, Shanghai, 200241, China
[*]Correspondence to: ychen@ce.ecnu.edu.cn

**ABSTRACT**
Integrating functions such as radar and communication into a single system is of great significance for the miniaturization and functional integration of future electronic warfare and 6G systems. Here, we show a dual-functional microwave photonic system for concurrent radar and secure communication. The scheme utilizes microwave photonic frequency multiplying and frequency conversion techniques to shift both the intermediate frequency radar and communication signals to the same frequency band, enabling radar and communication operations at the same time and frequency. The high-power radar signal is also used to mask the communication signal, increasing the difficulty of signal interception and thus enhancing security. By employing de-chirping at the radar receiver and self-interference cancelation at the communication receiver, the radar function can be implemented and the communication signal can also be correctly demodulated after removing the radar masking. An experiment is performed. A 0.3-GHz bandwidth linearly frequency-modulated signal is quadrupled and superimposed with two up-converted 0.5-Gbaud orthogonal frequency-division multiplexing signals. A communication data rate of 2 Gbit/s, a radar ranging measurement error of less than ±0.3 cm, and a radar inverse synthetic aperture radar imaging resolution of 12.5×10.2 cm are achieved.

**Keywords:** Concurrent radar and communication, co-time co-frequency, secure communication, self-interference cancelation, microwave photonics.

## 1. Introduction

Radar and secure communication play vital roles in modern electronic warfare, as they enable the active detection of targets and ensure the secure transmission of battlefield information without interception, respectively [1]. Integrating radar and communication functions facilitates hardware resource sharing, thereby reducing the system's volume and cost. Furthermore, it enables spectrum resource sharing, ultimately enhancing spectrum utilization efficiency. Besides, more sophisticated signal designs can reduce the risk of signals being intercepted [2]. Apart from the field of electronic warfare, the integration of radar and communication also holds promising application prospects in future 6G systems [3]. However, conventional electronic solutions face inherent electronic bottlenecks when applied to high-frequency and broadband application scenarios in future electronic warfare and 6G. As a result, they struggle to meet the application requirements of broadband tunable integrated radar and communication systems.

To address the above issues, dual-functional microwave photonic systems for concurrent radar and communication [4] have been widely investigated. Integrating radar and communication functions into a single microwave photonic system requires a balance to be struck in terms of hardware complexity, spectrum occupancy, and the performance of both functions, which requires careful design of both the waveforms and operating modes for radar and communication functions.

Hardware complexity for microwave photonic systems can be reduced by time-division multiplexing (TDM) [5] or frequency-division multiplexing (FDM) [6, 7] of radar and communication signals in the electrical domain. Recently, the time-frequency division multiplexing [8], which is a combination of TDM and FDM, is employed to allocate spectrum and time resources more flexibly according to the needs of radar and communication to improve the spectrum efficiency. However, the above multiplexed signals need to be generated in the electrical domain, which places extremely high demands on electronic systems when operating at very high frequencies and large bandwidths.

Sharing the waveform is another method for concurrent radar and communication operation. In this method, the shared waveform can be generated in the microwave photonic system, so the difficulty of generating multiplexed signals in the electrical domain can be resolved. The commonly used communication signals, such as the quadrature phase-shift keying (QPSK) signal [9] and orthogonal frequency-division multiplexing (OFDM) signal [10] can also be used for radar detection. Directly employing the aforementioned communication signals to achieve dual functionality is not only straightforward but also avoids the use of complex waveforms. However, in this scenario, radar receivers necessitate higher sampling rates and increased processing complexity. Based on linearly frequency-modulated (LFM) signals, which can undergo de-chirping processing to significantly reduce the complexity of receiving and processing at the radar receiver, additional communication data is loaded onto the LFM signal to generate a shared waveform for concurrent radar and communication systems, such as amplitude-shift keying LFM (ASK-LFM) signal [11,12], PSK-LFM signal [13,14], and constant envelope LFM-OFDM signal [15,16]. However, the communication data rate of the shared waveform based on LFM signals is generally much lower than the bandwidth that their occupied frequency band can support. Additionally, the communication data loaded onto the LFM signals can greatly affect radar performance [13].

From the above, it can be seen that it is difficult for existing concurrent radar and communication systems to simultaneously achieve low hardware complexity, high spectrum utilization, and good performance for both functions. Besides, in electronic warfare, the communication security of concurrent radar and communication systems is also a crucial issue when confronted with various electronic interception methods. This problem has not been addressed in the reported microwave photonic concurrent radar and communication systems.

In this paper, a dual-functional microwave photonic system for concurrent radar and secure communication is proposed. This scheme utilizes microwave photonic frequency multiplying and frequency conversion techniques to shift both the intermediate frequency (IF) LFM signal and the IF QPSK-OFDM signal to the same frequency band, enabling concurrent operation of communication and radar functions at the same time and frequency. The communication signal, which has a lower power, is masked by the greater power of the radar signal to ensure secure data transmission. The radar signal masking can be removed at the receiver using the self-interference cancellation (SIC) technique. If the signal is intercepted, the adversary cannot receive the communication signal due to their ignorance of the radar signal format. An experiment is performed. A 0.3-GHz bandwidth LFM signal undergoes a quadrupling process in both frequency and bandwidth to 13.5 to 14.7 GHz and is then superimposed with two up-converted QPSK-OFDM signals, each operating at 0.5 Gbaud. A communication data rate of 2 Gbit/s, a radar ranging measurement error of less than ±0.3 cm, and a radar inverse synthetic aperture radar (ISAR) imaging resolution of 12.5×10.2 cm can be simultaneously achieved.

## 2. Principle and experimental results

The schematic of the proposed dual-functional microwave photonic system is shown in Fig. 1. A continuous-wave light wave with a frequency of 193.4098 THz and a power of 13.5 dBm is generated by a laser diode LD (ID Photonics CoBriteDX1-1-C-H01-FA), which is sent to a dual-polarization quadrature phase-shift keying (DP-QPSK) modulator (Fujitsu FTM7977HAQ/331). An IF-LFM signal with an amplitude of 400 mV, a center frequency of 3.525 GHz, and a bandwidth of 0.3 GHz is generated from an arbitrary waveform generator (AWG1, Keysight M8195A, 64 GSa/s), amplified by an electrical amplifier (EA1, Multilink MTC5515), and then applied to one RF port of the X-polarization dual-parallel Mach–Zehnder modulator (DP-MZM) in the DP-QPSK modulator. The other RF port of this DP-MZM is with no input. The X-polarization DP-MZM is biased to generate the ±2nd-order optical LFM sidebands with carrier suppressed. Two IF QPSK-OFDM signals with center frequencies of 1.8 and 2.4 GHz and the same baud rate of 0.5 Gbaud are also generated by the AWG. The reason for selecting two OFDM signals here is to simulate the scenario where there are two different communication users. The two IF QPSK-OFDM signals are amplified by EA2 (Teledyne cougar AR3069B) and then sent to one input port of a 90° hybrid coupler (90° HYB). The amplitude of the OFDM signal from the AWG is set to 150, 100, 75, or 0 mV in the experiment. A local oscillator (LO) signal with a frequency of 12.1 GHz and a power of 10 dBm is generated from a microwave signal generator (MSG, Agilent 83630B) and then divided by a 3-dB electrical coupler (EC1, Narda 4456-2 2–18 GHz). One output of EC1 is injected into another input port of the 90° HYB. The two outputs of the 90° HYB are applied to the two RF ports of the Y-polarization DP-MZM in the DP-QPSK modulator. By biasing this DP-MZM as a carrier-suppressed single-sideband modulator, a pair of opposite first-order optical sidebands, originating respectively from the OFDM signal and the LO signal, are output by the Y-polarization DP-MZM. Then, the optical signals from the DP-QPSK modulator, as shown in Fig. 1(a), are detected in a photodetector (PD, u2t MPRV1331A).

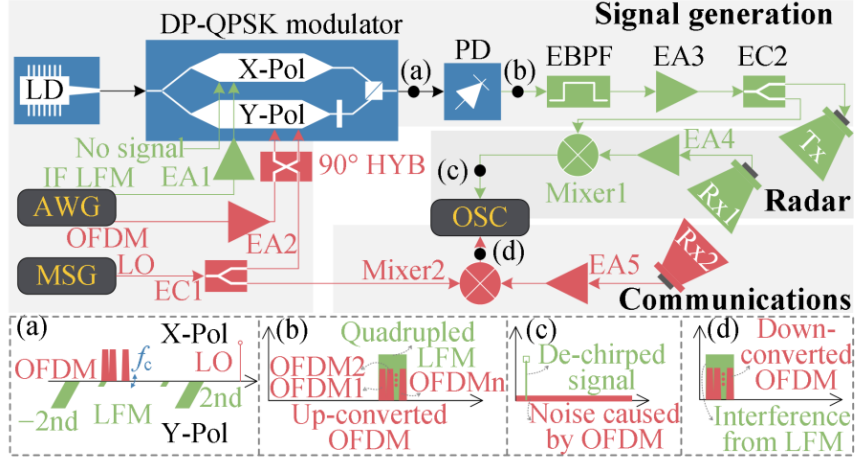

Fig. 1. Schematic of the proposed dual-functional microwave photonic system. LD, laser diode; DP-QPSK, dual-polarization quadrature phase-shift keying; AWG, arbitrary waveform generator; IF, intermediate frequency; LFM, linearly frequency-modulated; OFDM, orthogonal frequency-division multiplexing; LO, local oscillator; EA, electrical amplifier; MSG, microwave signal generator; EC, electrical coupler; 90° HYB, 90° hybrid coupler; PD, photodetector; EBPF, electrical bandpass filter, Tx, transmitting antenna; Rx, receiving antenna; OSC, oscilloscope; Pol, polarization. (a)-(d) are the schematic spectrum diagrams of the signals at different locations marked in the system diagram.

Due to the orthogonal polarization of the optical signals output by the two DP-MZMs, they will be independently converted to the electrical domain in the PD and then coupled and output from the RF port of the PD. The generated frequency-and-bandwidth-quadrupled LFM signal and up-converted OFDM signal are schematically shown in Fig. 1(b), which occupy the same frequency range and appear simultaneously in the time domain. The co-time co-frequency signal from the PD is filtered by an electrical bandpass filter (EBPF, KGL YA357-2, 13.4–17.12 GHz), amplified by EA3 (CLM145-7039-293B, 5.85–14.50 GHz, 39 dB), and then equally divided into two parts by EC2 (MCLI PS2-11, 2–18 GHz).

One output of EC2 is radiated to the free space by using a transmitting antenna (Tx) and the other output of EC2 is injected into the LO port of Mixer1 (M/A-COM M14A) at the radar receiver. The radar echo signal which is reflected by the targets is received by one receiving antenna (Rx1), amplified by EA4 (CLM145-7039-293B, 5.85–14.50 GHz, 39 dB), and finally injected into the RF port of Mixer1. The de-chirped signal from Mixer1 is captured by a real-time oscilloscope (OSC, R&S RTO2032) and further processed in the digital domain for radar ranging and imaging. Although the OFDM signal is mixed with the LFM signal, after de-chirping at Mixer1, the LFM signal is compressed into frequency-domain pulses and the OFDM signal will not. Since the power of the LFM signal is much greater than that of the OFDM signal, the OFDM signal can be regarded as noise. Therefore, the OFDM signal mixed with the LFM signal mainly affects the noise floor of the signal after de-chirping.

Rx2 at the communication receiving end is 0.6 m away from the Tx in the experiment. The signal received by Rx2 is amplified by EA5 (ALM 145-5023-293 5.85–14.5 GHz, 23 dB) and then injected into the RF port of Mixer2 (M/A-COM M14A), whereas the signal from the other output of EC1 is injected into the LO port of Mixer2. In practical systems, the LO signal here can be directly generated at the communication receiver, without the need to be distributed from the transmitter. After mixing in Mixer2, the down-converted received signal is obtained from the IF port of Mixer2. The IF signal not only contains the desired communication signal but also includes a high-power LFM signal. In practice, the communication receiver needs to have complete knowledge of the radar signal to eliminate it from the IF signal through further processing in both the analog domain and digital domain. The knowledge of radar signals can be predetermined between the transmitter and the communication receiver or transmitted through other means. To simplify the system and for proof of concept, we have omitted the analog cancellation part. Instead, after acquiring the IF signal using the OSC, we reconstruct and cancel the radar signal only in the digital domain [17]. After removing the LFM interference, the OFDM information can be recovered successfully.

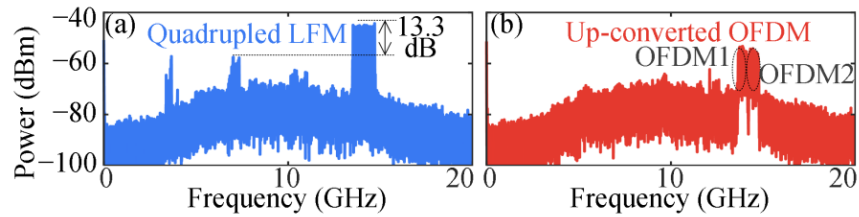

Fig. 2. Electrical spectra of the generated (a) frequency-and-bandwidth-quadrupled LFM signal and (b) up-converted OFDM.

Figures 2(a) and (b) show the spectra of the generated frequency-and-bandwidth-quadrupled LFM signal and up-converted OFDM signal when the amplitude of the IF OFDM signal is 150 mV, observed by an electrical spectrum analyzer (Ceyear 4052E). The uneven noise floor is introduced by an electrical amplifier because a directional coupler (EC3, KRYTAR MODEL 1818, 2–18 GHz, −16 dB) and EA6 (CLM145-7039-293B, 5.85–14.50 GHz, 39 dB) are employed when observing the spectrum.

When focusing on communication functions, the LFM signal is considered as an interference to the OFDM signal, and the signal-to-interference ratio (SIR) is defined as the power ratio between the OFDM signal and the LFM signal. Figure 3(a) shows the electrical spectra of the down-converted LFM and OFDM signals and the constellation diagrams with and without SIC when the SIR is −7.8 dB. The electrical spectra are obtained by performing a fast Fourier transform (FFT) to the waveform captured by the OSC. When the SIC is not employed, the LFM interference completely overwhelms the OFDM signal, making it difficult to directly demodulate the OFDM signal, as shown in the constellation diagrams in Figs. 3(a-i) and (a-ii). When the LFM interference is suppressed by the digital SIC, an interference cancellation depth of 21.3 dB is achieved. As can be seen from Fig. 3(a), the two OFDM signals for two users can be distinguished from the spectrum. Figures 3(a-iii) and (a-iv) show the corresponding constellation diagrams with four distinguished constellation points. For comparison, the electrical spectrum of the OFDM signals when the LFM signal is not applied is given in Fig. 3(b), while the corresponding constellation diagrams are given in Figs. 3(b-i) and (b-ii). As can be seen, the spectrum and constellation diagrams are very similar to those with SIC given in Fig. 3(a). Figure 3(c) gives the error vector magnitude (EVM) of two OFDM signals when the LFM signal is not applied and when the LFM interference is applied and SIC is performed. As the SIR decreases, the EVM deteriorates. In the presence of radar LFM signal masking, even after SIC is applied, the EVM is still inferior to that in the absence of LFM interference. This is primarily due to the fact that SIC cannot completely eliminate the LFM interference, and the residual LFM interference, acting as noise, causes a certain degree of performance degradation. However, at the cost of a slight performance degradation, LFM masking can be achieved, thereby enhancing the anti-interception capability of the communication.

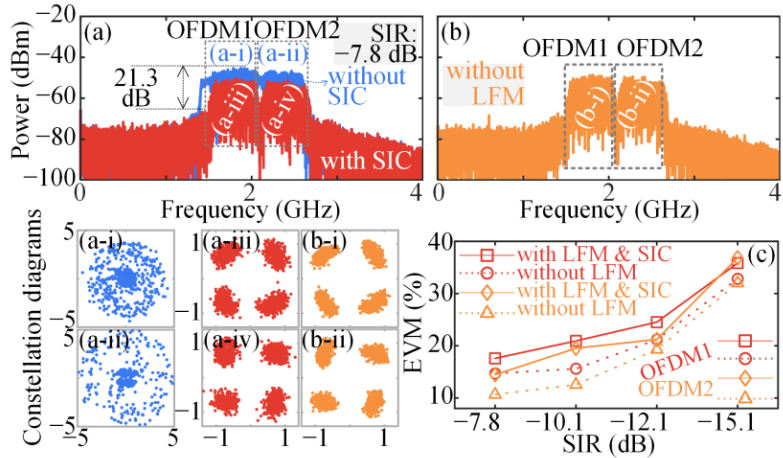

Fig. 3. (a) Electrical spectra and (a-i) to (a-iv) constellation diagrams of the signal with and without SIC when the SIR is −7.8 dB. (b) Electrical spectrum and (b-i) and (b-ii) constellation diagrams of the signal when only the OFDM signal is applied. (c) EVM of two different OFDM signals under different SIRs.

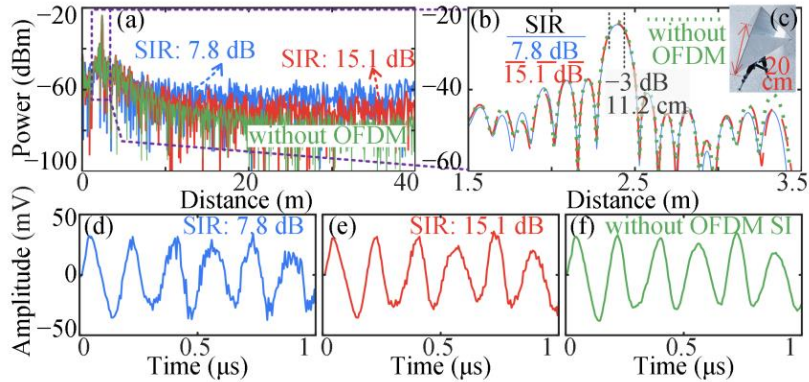

Fig. 4. Ranging results of one corner reflector at different SIRs. (a) Electrical spectra. (b) Zoomed-in view of the purple-dotted rectangular area in (a). (c) Photograph of the corner reflector. (d)-(f) Waveforms corresponding to (a).

Then the radar function is investigated. In this case, the OFDM signal is the interference, so the SIR is defined as the power ratio between the LFM signal and the OFDM signal. Figure 4 shows the radar ranging results when a corner reflector shown in Fig. 4(c) is used as the target. The de-chirped signal at different SIRs is shown in Fig. 4(a) while a zoomed-in view of the peaks is given in Fig. 4(b). It is observed that the range resolution is around 11.2 cm, while the theoretical value is 12.5 cm. It can also be observed that, in this case, the de-chirped signals exhibit very similar spectra near their peaks. The primary difference in the spectrum lies in the noise floor: As the OFDM interference increases (SIR decreases), the noise floor is elevated. Figures 4(d)–(f) show the waveforms from 0 to 1 μs corresponding to the spectra in Fig. 4(a). It can be seen that the lower the SIR, the greater the noise in the waveform. Although the OFDM signal can coexist with the LFM signal at the same time and frequency, they elevate the noise floor of the system, thereby reducing the system's ability to detect weak signals.

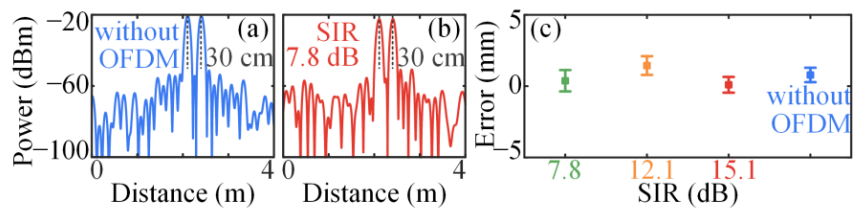

Fig. 5. Ranging results of two targets. Spectra of the de-chirped signals (a) without and (b) with the OFDM interference. (c) Measurement error of the distance between two targets in 40 measurements.

Figure 5 shows the ranging results of two corner reflectors separated by 30 cm. Figures 5(a) and (b) show the spectra of the de-chirped signals when the OFDM interference signal is not applied and when the SIR is 7.8 dB, respectively. The distance of the two targets can be obtained by measuring the distance of the two peaks. The distance measurement error of the two targets under different SIRs is shown in Fig. 5(c). When the SIR increases from 7.8 dB to 15.1 dB and further to infinity (without OFDM signal), the measurement error is smaller than ±0.3 cm in 40 measurements, and the standard deviation of the error decreases from 0.75 to 0.55 mm. The decrease in standard deviation as the SIR increases is mainly due to the reduction in OFDM interference when the SIR rises. This leads to a decrease in the noise introduced to the radar de-chirped signal by the OFDM interference, thereby enhancing the consistency of the measurement results in different measurements.

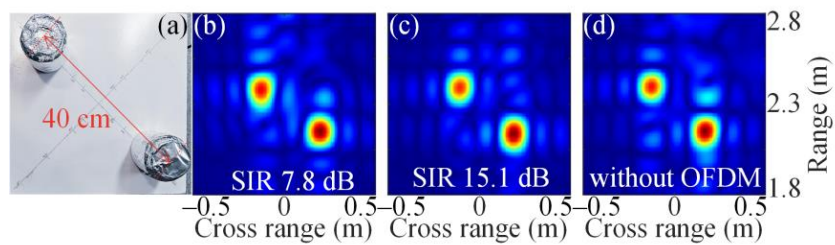

Fig. 6. (a) Photograph of two cylinders. (b)–(d) ISAR imaging results under different SIRs.

Finally, ISAR imaging is demonstrated. Two cylinders, each having a diameter of 10.5 cm and a height of 14.5 cm, are positioned on a turntable that rotates with a period of 24.56 s. The accumulation time in ISAR imaging is set to 0.4 s, so the theoretical imaging resolution is 12.5 × 10.2 cm. Fig. 6(a) shows the photograph of the two cylinders. Figures 6(b)–(d) show the ISAR imaging results at different SIRs. We can observe that, under conditions of strong echoes in Fig. 6, the imaging results obtained at different SIRs are very close within the given coordinate range, which is consistent with the conclusions drawn from Fig. 4. During the imaging experiment, due to the requirement for 0.4 s of data accumulation, limitations in the sampling rate and oscilloscope storage depth prevented us from displaying the noise floor at relatively greater distances in Fig. 4.

If such information could be displayed, it would be evident that there are differences in the background noise under different SIRs.

## 3. Conclusion

In summary, a dual-functional microwave photonic system for concurrent radar and secure communication is proposed and demonstrated. The key contribution of this work lies in our use of high-power radar signals to mask communication signals for co-time co-frequency radar and communication. By doing so, we not only fulfill the radar functions but also significantly increase the difficulty of intercepting the communication signals. Furthermore, by combining the SIC technique, the communication receiver can correctly receive the signals. The concept is demonstrated by an experiment. Microwave photonic frequency multiplying and conversion are employed for the co-time co-frequency radar and communication signal generation at 13.5–14.7 GHz. A communication data rate of 2 Gbit/s, a radar ranging measurement error of less than ±0.3 cm, and a radar ISAR imaging resolution of 12.5×10.2 cm can be simultaneously achieved. This research is of great significance for the integrated implementation of different functions and the miniaturization of systems in future electronic warfare and 6G systems.

**Acknowledgements**

National Natural Science Foundation of China (62401207, 61971193); Key Laboratory of Radar Imaging and Microwave Photonics, Nanjing University of Aeronautics and Astronautics, Ministry of Education (NJ20240004); Science and Technology Commission of Shanghai Municipality (22DZ2229004).